\documentclass[pre,twocolumn,preprintnumbers,amsmath,amssymb,superscriptaddress,floatfix]{revtex4}

\usepackage{graphicx}
\usepackage{dcolumn}
\usepackage{bm}
\usepackage{color} 
\usepackage{amsfonts}
\usepackage{amsmath}
\usepackage{amsthm}
\usepackage{lipsum}
\usepackage{array}
\usepackage[integrals]{wasysym}

\renewcommand{\vec}[1]{\mathbf{#1}}

\begin{document} 
 
\title{Swelling of ionic microgel particles in the presence of excluded-volume interactions: a density functional approach}

\author{A.\ Moncho-Jord\'{a}\footnote{Corresponding author: moncho@ugr.es}} 
\affiliation{Departamento de F\'{\i}sica Aplicada, Facultad de Ciencias, Universidad de Granada, Campus Fuentenueva S/N, 18071 Granada, Spain.}
\affiliation{Instituto Carlos I de F\'{\i}sica Te\'{o}rica y Computacional, Facultad de Ciencias, Universidad de Granada, Campus Fuentenueva S/N, 18071 Granada, Spain.}

\author{J. Dzubiella} 
\affiliation{Institut f\"ur Weiche Materie und Funktionale Materialien, Helmholtz-Zentrum Berlin, Hahn-Meitner-Platz 1,14109 Berlin, Germany.}
\affiliation{Institut f\"ur Physik, Humboldt-Universität zu Berlin, Newtonstr. 15, 12489 Berlin, Germany.}
\affiliation{Multifunctional Biomaterials for Medicine, Helmholtz Virtual Institute, 14513 Teltow, Germany.}
 
\begin{abstract}
In this work a new density functional theory framework is developed to predict the salt-concentration dependent swelling state of charged microgels and the local concentration of monovalent ions inside and outside the microgel. For this purpose, elastic, solvent-induced and electrostatic contributions to the microgel free energy are considered together with the free energy of the ions. In addition to the electrostatic interaction, the model explicitly considers both the microgel-ion excluded-volume (steric) repulsion and the ionic correlations, in such a way that the formalism is consistent with the Hypernetted-Chain Closure approximation (HNC). We explore the role that the solvent quality, chain elasticity, salt concentration and microgel bare charge play on the swelling state, the effective charge and on the ionic density profiles. Our results show that the microgel-ion steric exclusion foments the increase of the particle size up to 10\%. The role that the steric effect plays on the counterion distribution becomes more important when the microgel approaches the shrunken configuration, developing an accumulation peak at the microgel interface and a reduction in the inner core of the microgel that induce a significant increase of the microgel effective charge. We further find that deep inside the particle charge electroneutrality is achieved and a Donnan potential corrected by the steric exclusion is established.
\end{abstract} 
 
\maketitle 

\section{Introduction}
\label{introduction}

A microgel (or nanogel) particle is formed by a cross-linked polymer network of colloidal size immersed in a solvent, which can be designed to swell or shrink in response to many external parameters, such as temperature, pH, and solvent quality among others~\cite{Murr95,Saun99,Fern11,Zhou15}. Particles formed by the copolymerization of monomers of $N$-isopropylacrylamide (PNIPAM) or $N$-vinylcaprolactam (PVCL) are two examples of microgels. Due to their nanometric size, the timescale of the swelling response (which is roughly proportional to the square of the typical spatial dimension of the microgel) is of the order of seconds, which is very short compared to the ones observed in the so-called macroscopic gels~\cite{Tana79,Long14}. Furthermore, the soft and porous nature of the microgels allow them to be permeated by the solvent, ions and other neutral or charged macromolecules. The combination of these properties make microgel suspensions unique smart materials for industrial and biomedical applications, such as carrier particles for biomolecules or controlled drug release~\cite{Ramo11,Lesh,Kenn16}.

The encapsulation of solutes inside microgels may depend on many parameters, such as the swelling state of the microgel, its internal distribution of bare charge, the net charge of the solute, or the hydrophobic character of both the solute and the polymer network~\cite{Lope04}. In the case of charged solutes immersed in a suspension of ionic microgels, the total amount of absorbed molecules may be also strongly influenced by the presence of ions. In fact, the local concentration of counter- and coions inside and around the microgel may play a determinant role on how the sorption of solute takes place (externally at the surface, i.e., adsorption, or deep inside the polymer network, i.e, absorption).

Due to the local variation of the ionic density profiles close to the particle surface, charge electroneutrality is not fulfilled at every point inside the particle. In this respect, different theoretical approaches based on integral equations and linear response theory clearly indicate that the ionic distribution leads to a non-zero effective (or net) charge inside the microgel, $Z_{eff}$~\cite{Dent03,Gott05,Monc13a}. Moreover, $Z_{eff}$ is significantly altered when ion-specific effects are taken into account in addition to electrostatic ones. For instance, its absolute value grows when counterions become expelled from the internal volume of the particle due to the excluded-volume repulsion exerted by the polymer mesh~\cite{Monc13b}. Oppositely, $| Z_{eff} |$ may decrease and even show charge inversion when counterions are specifically attracted to the polymer network by means of short-range hydrophobic forces~\cite{Monc14}. In all these works, however, the microgel was represented by a fixed object which creates a constant external field for the ions, but the swelling response of the microgel was not accounted for. 

The swelling behavior of ionic microgel particles were formerly tackled on a coarse-grained level by means of a combination of the Flory-Huggins theory for the elastic and solvent-induced contributions, and the Debye-H\"{u}ckel linear screening approximation for the effect of the counterions~\cite{Levi02,Lope04,Fern00,Hoar07}. Some recent works address the microgel volume transition in the presence of ions or charged surfactants using more sophisticated theories and simulation methods~\cite{Koso15,Rumy15,Rumy14}. However, all those approaches did not consider ionic correlations nor excluded-volume effects. Sing {\em et al.}~\cite{Sing13} applied an integral equation formalism that considers the finite-size ionic correlations to predict a reentrant swelling for large salt concentrations. For this purpose, they used a two-phase description (the macroscopic gel and the bulk phase) so that the exchange of ions between both phases was permitted at a fixed chemical potential and preserving electroneutrality. A similar grand canonical description was employed by Ahualli {et al.}~\cite{Ahua14} but in this case making use of coarse-grained computer simulations to model the macroscopic gel, also under electroneutral conditions. They compared the simulation results for microgel swelling with a theoretical approach based on the Poisson-Boltzmann equation, and found good agreement as soon as excluded-volume effects promoted by the cross-linked polymer matrix were properly considered. Recently, Colla {et al.}~\cite{Colla14} went beyond this two-phase description and studied the uniform swelling and the local variation of the ionic concentrations for finite-sized and uniformly charged microgel particles. In that work, the authors employed a density functional approach that included the free energy of both the microgel particle and the ions. Their model, however, is a mean-field Poisson-Boltzmann approach that neglects the ion correlations and the excluded-volume effects, which have been proven to be important for de-swollen configurations~\cite{Monc13b,Adro15}.

The main goal of this work is to propose a density functional theory (DFT) able to predict consistently both the equilibrium counterion and coion density profiles ($\rho_+(r)$ and $\rho_-(r)$, respectively) and the swelling response, but taking special care of considering ion-specific excluded-volume effects and finite-size ion correlations in addition to the electrostatic interactions. The model gathers the elastic, solvent and electrostatic free energy contributions coming from the polymer network inside the microgel, together with the free energy of the ions in the presence of the microgel. Moreover, the ion free energy term is built to be compatible with Ornstein-Zernike (OZ) integral equation theory within the Hypernetted-Chain Closure (HNC) for ion-ion and microgel-ion correlations. Although this approximation neglects the bridge functions, it has been shown to perform quite well when compared to other theories and simulations~\cite{Monc13a,Adro15}. For this purpose, a quadratic functional Taylor expansion with respect to the bulk densities of the ions is included in the ionic free energy, which accounts for the ionic correlation beyond the mean-field electrostatic treatment. Similar approximations have been used in the literature to describe various charged soft matter systems such as grafted polyelectrolytes~\cite{Jian07,Jian08}, polyelectrolytes near oppositely charged interfaces~\cite{Li06}, ions near charged electrodes~\cite{Alts87}, and mixtures of charged macroions or colloids and electrolyte~\cite{Li04,Yu04}. We focus on the particular case of 1:1 electrolyte suspensions, although the generalization of this method to multivalent ions is straightforward.

The paper is organized as follows: In Section~\ref{system} the particle interactions among the different components of the system (microgels, counterions and coions) are described. Section~\ref{OZHNC} briefly explains the OZ-integral equations method and the HNC relation used to determine the ionic density profiles around the charged microgel. Then, a DFT consistent with HNC that includes both ionic correlations and the microgel-ion excluded-volume interactions, is developed in Section~\ref{DFT} to incorporate also the microgel swelling. Section~\ref{implementation} specifies the system conditions and describes the details related to the numerical implementation of the method. In Section~\ref{results} the theoretical predictions for the particle swelling, effective charge and local ionic concentrations are shown in terms of many system parameters. Finally, we summarize the most important results in Section~\ref{conclusions}.

\section{The model system}
\label{system}

We consider a three-component mixture formed by microgels, counterions and coions (indexes $m$, $+$ and $-$, respectively) immersed in a continuous solvent with a electric permittivity, $\epsilon$.  Counterions (coions) are assumed to be modeled by charged hard spheres of radius $R_+$ ($R_{-}$) and valence $Z_+$ ($Z_{-}$). This allows to account for the finite size effects. Microgels are here treated as permeable spheres with a uniform mass holding a total bare charge $Z_m$ homogeneously distributed within a sphere of radius $R_m$. It should be emphasized that $R_m$ is not a constant quantity, since it depends on many parameters such as temperature, salt concentration, microgel charge, etc. The polymer volume fraction inside the microgel particle may be written as
\begin{equation}
\label{phi}
\phi = \phi_0 \left(R_{0}/R_m \right)^3
\end{equation}
where $R_{0}$ and $\phi_0$ are the radius and polymer packing fraction of the microgel in a reference swelling state.

In order to determine the equilibrium ionic density profiles inside and around the microgel particle we need to know the analytic expression of the pair interaction potentials. The dimensionless pair potentials between ions are given by 
\begin{equation}
\label{vij}
\beta V_{ij}(r)=\left\{\begin{array}{l@{\quad}l}
\infty & r\leq R_{i}+R_{j} \\
Z_iZ_j l_B/r & r > R_{i}+R_{j}
\end{array}  \right. 
\end{equation}
where $r$ is the distance between the centers of both ions, $i,j=\pm$, and $l_B$ is the Bjerrum length, defined as $l_B=e^2/(4\pi \epsilon k_B T)$, where $k_B$ is the Boltzmann constant, $T$ the absolute temperature and $\beta =1/(k_BT)$.

\begin{figure*}
 \centering
 \includegraphics[height=6cm]{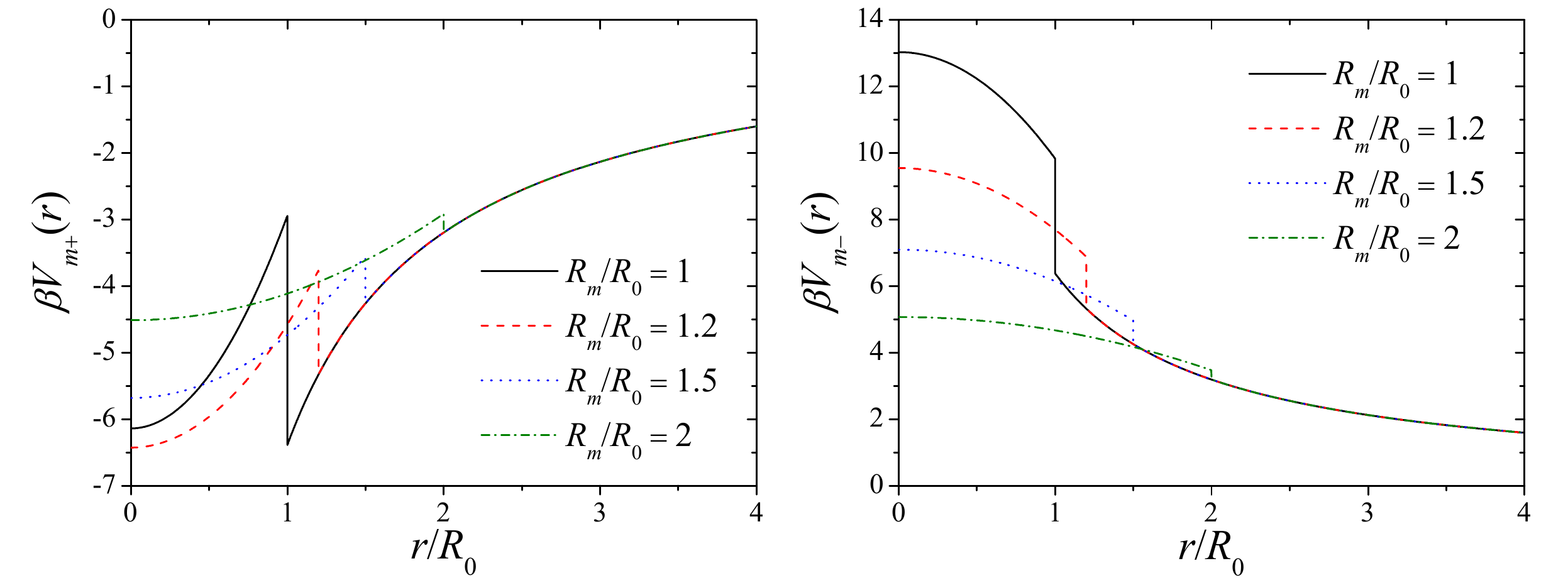}
 \caption{Bare interactions between the microgel and a monovalent counterion (left panel) or coion (right panel) for different swelling ratios, $R_m/R_0$. These pair potentials are obtained from eqns~(\ref{vmi}), (\ref{vmielec}) and (\ref{vmister}) assuming $R_0=30$~nm, $\phi_0=0.64$, $Z_m=-270$, $R_+=R_-=0.2$~nm and $R_{mon}=0.4$~nm.}
 \label{bare_potentials}
\end{figure*}
The pair interaction between ions and a microgel particle will be split into an electrostatic and excluded-volume additive contributions~\cite{Monc13b}
\begin{equation}
\label{vmi}
V_{mi}(r) = V_{mi}^{elec}(r) + V_{mi}^{exc}(r) \ \ \ \ \ i=\pm.
\end{equation}
The first term of eqn~(\ref{vmi}) corresponds to the electrostatic potential energy between a single ion and the microgel. For a uniformly charged spherical microgel of radius $R_m$, this contribution is given by 
\begin{equation}
\label{vmielec}
\beta V_{mi}^{elec}(r)=\left\{\begin{array}{l@{\quad}l}
\frac{Z_mZ_il_B}{2R_m}\left( 3-\frac{r^2}{R_m^2} \right) & r\leq R_m \\
\frac{Z_mZ_il_B}{r} & r > R_m.
\end{array}  \right. 
\end{equation}
Here, $r$ is represents the distance between the microgel and ion centers. The second term of eqn.~\ref{vmi} accounts for the ion-specific excluded-volume repulsion that an incoming ion experiences when diffusing inside the polymer network. For point-like ions this interaction is only dependent on the free volume left by the polymer fibers. However, for finite size ions, this repulsion depends also explicitly on the radius of the monomeric units ($R_{mon}$), the ion size, and on the internal morphology of the cross-linker polymer network. One of the first models to account for internal structure of gels assumed that they could be represent by interconnected spherical pores~\cite{Fati03}. This hypothesis, however, did not succeed to predict the simulation results obtained for the partition coefficient of neutral solutes inside cross-linked polymer networks~\cite{Ques14a}. Here, we assume that the polymer network is roughly given by an assembly of randomly placed spherical monomers. This approximation has been successfully employed to predict the ionic density profiles inside and outside a single microgel and the permeation of neutral and charged solutes obtained via Monte Carlo simulations~\cite{Ahua14,Adro15}. Under this assumption, the ion-microgel steric (excluded-volume) repulsion may be analytically calculated~\cite{Lazz00}
\begin{equation}
\label{vmister}
\beta V_{mi}^{exc}(r)=\left\{\begin{array}{l@{\quad}l}
-\ln \left( 1-\phi \right)\left( 1 + R_i/R_{mon} \right)^3 & r\leq R_m \\
0 & r > R_m
\end{array}  \right. 
\end{equation}
Figure~\ref{bare_potentials} depicts the counterion-microgel and coion-microgel bare interactions ($V_{m+}(r)$ and $V_{m-}(r)$, respectively) for a particular case. The pair potentials show a Coulombic decay for distances $r>R_m$. For $r<R_m$ the potential is soft and reaches a maximum/minimum at the center of the microgel, $r=0$, where the electric field created by the particle is zero. As it may be observed, the steric repulsion introduces a repulsive barrier located at $r=R_m$ that partially hinders the ionic permeation inside the microgel. The barrier height grows with the polymer volume fraction, $\phi$, and so it plays a more important role in shrunken states, whereas it only represents a minor perturbation for swollen conformations. Please note that any discontinuity in the pair potential (given by the steric barrier $\Delta V = V_{mi}^{exc}(r<R_m)$), yields also a discontinuity in the ionic concentrations, so that $\rho_{i}(R_m^+)=\exp (\beta \Delta V)\rho_{i}(R_m^-)$. Therefore, the jump of the ionic densities grows exponentially with $\Delta V$, leading to huge peaks in the density profiles of counterions at $r=R_m$ for shrunken conformations.

\section{HNC-Ornstein-Zernike equations}
\label{OZHNC}

The main aim of the work is to develop a DFT consistent with Ornstein-Zernike (OZ) integral equation theory  within the HNC approximation. In this Section we briefly discuss how the integral equations method are able to provide the ionic density profiles around a fixed microgel particle. In the limit of very dilute microgel suspensions, these equations are given (in the Fourier space) by~\cite{Hans06}
\begin{eqnarray}
\left. \begin{array}{l@{\quad}l}\label{OZ1}
\hat{h}_{++} = \hat{c}_{++} + \rho_{+}^b \hat{c}_{++}\hat{h}_{++} + \rho_{-}^b \hat{c}_{+-}\hat{h}_{+-} &   \\
\hat{h}_{+-} = \hat{c}_{+-} + \rho_{+}^b \hat{c}_{++}\hat{h}_{+-} + \rho_{-}^b \hat{c}_{+-}\hat{h}_{--} & \\
\hat{h}_{--} = \hat{c}_{--} + \rho_{+}^b \hat{c}_{+-}\hat{h}_{+-} + \rho_{-}^b \hat{c}_{--}\hat{h}_{--} &
\end{array} \right\}, 
\end{eqnarray}
for the ion-ion correlations, and
\begin{eqnarray}
\left. \begin{array}{l@{\quad}l}\label{OZ2}
\hat{h}_{m+} = \hat{c}_{m+} + \rho_{+}^b \hat{c}_{++}\hat{h}_{m+} + \rho_{-}^b \hat{c}_{+-}\hat{h}_{m-} &  \\
\hat{h}_{m-} = \hat{c}_{m-} + \rho_{+}^b \hat{c}_{+-}\hat{h}_{m+} + \rho_{-}^b \hat{c}_{--}\hat{h}_{m-} &  
\end{array} \right\}, 
\end{eqnarray}
for the ion-microgel correlation. $\rho_+^b$ and $\rho_-^b$ represent the number density of counterions and coions in the bulk, far away from the perturbation caused by the microgel particle. These equations consider the particular case of a single microgel particle, and so, they actually correspond to an infinitely diluted microgel suspension, $\rho_m^b\rightarrow 0$ (the study of the ionic density profiles for concentrated colloidal suspensions would necessarily require the knowledge of the microgel-microgel interaction potential). $h_{ij}(r)$ and $c_{ij}(r)$ are the so-called total and direct correlation functions. In order to solve these equations, five additional closure relations are required to couple both functions. In this work, the Hypernetted-Chain Closure (HNC) is used for all particle correlations
\begin{equation}
\label{HNC}
h_{ij}(r) = \exp \left[ - \beta V_{ij}(r) + h_{ij}(r) - c_{ij}(r) \right]  - 1,
\end{equation}
as it has shown to represent a quite accurate approximation for ionic microgels in salty suspensions~\cite{Monc13a,Monc13b}.
Using the bulk ionic concentrations and the pair interaction potentials (see eqns~(\ref{vij}), (\ref{vmi}), (\ref{vmielec}) and (\ref{vmister})) as input functions, both sets of equations are successively iterated until convergence is achieved. The sought ionic density profiles are finally given by
\begin{equation}
\rho_{i}(r)=\rho_i^b\left[ h_{mi}(r)+1 \right] \ \ \ {\rm with}\ \ i=\pm.
\end{equation}
It should be noted that the electroneutrality condition
\begin{equation}
\label{electroneutrality}
Z_+\rho_+^b+Z_-\rho_-^b=0
\end{equation}
is automatically satisfied when solving the OZ-HNC equations. This means that, as soon as we introduce in the system a charged microgel, there is an exchange of ions with the bulk reservoir in order to counterbalance the microgel charge. Integrating those ionic density profiles over the volume of the microgel yields the effective charge of the particle 
\begin{equation}
\label{effective_charge}
Z_{eff}=Z_m+4\pi\int_0^{R_m}\left[Z_+\rho_{+}(r)+Z_{-}\rho_{-}(r) \right] r^2dr .
\end{equation}

\section{Density functional theory}
\label{DFT}

In our model, the system is consistent of a large spherical open cell with a single charged microgel placed in the center, at $r=0$. Since we are interested in the knowledge of the ionic density profiles around a microgel particle, our functional must consider the microgel as an external potential for the ions. In particular, our external potential, $V_{mi}(r)$, is the one given by eqns~(\ref{vmi}), (\ref{vmielec}) and (\ref{vmister}). In the presence of the microgel, the ionic densities become non-uniform and show a dependence on the distance to the microgel center, $\rho_i(r)$. This heterogeneous system formed by the central microgel and the surrounding ions is in equilibrium with a homogeneous reservoir filled exclusively by ions: the bulk. The bulk number densities ${\rho_i^b}$ are determined from the salt concentration added to the microgel suspension. Finally, the mass equilibrium between the system and the reservoir is controlled by the chemical potential of counter- and coions, $\mu_{\pm}$.

In addition, microgel particles are deformable objects that can adopt different swelling states depending on the solvent conditions, the cross-linker concentration and the bare charge. In the presence of ions, the microgel interacts with them, and the equilibrium radius $R_m$ becomes also dependent of the salt concentration. Therefore, the grand canonical potential of the system should incorporate the free energy of the microgel, $F_m$, the free energy of the ions, $F_{ions}$ and the contribution coming from the microgel-ion interactions. Hence, we have
\begin{equation}
\label{omega}
\Omega [\{\rho_i(\vec{r})\},R_m] = F_m + F_{ions} + \sum_{i=\pm}\int \left[ V_{mi}(\vec{r})-\mu_i \right] \rho_i(\vec{r})d\vec{r}.
\end{equation}

\subsection{Free energy of the microgel}

The free energy of the microgel will be split into three different terms: elastic, solvent-induced, and electrostatic self-contributions
\begin{equation}
F_m = F_m^{elastic} + F_m^{solvent} + F_m^{self-el}
\end{equation}
For the elastic free energy we make use of the popular rubber elasticity model~\cite{Fern11}
\begin{equation}
\beta F_{elastic}=\frac{3N_{chains}}{2}\left[ \left(\frac{R_m}{R_0}\right)^2 - \ln\left( \frac{R_m}{R_0}\right) -1 \right]
\end{equation}
where $R_m$ is the microgel radius, $R_0$ is its radius in the undeformed state and $N_{chains}$ is the number of cross-linked chains. The elasticity strongly depends on the average chain length. We will define $\nu$ as the number of monomeric units per chain. The internal morphology of the microgel and the elastic response is specified by $\nu$. Indeed, short chains imply that the particle has a larger cross-linker concentration and so it is more difficult to stretch. On the contrary, for long chains the microgel is likely to be deformed by any external stimuli with a relatively small elastic free energy cost. Assuming that monomers have a spherical shape of radius $R_{mon}$, we have
\begin{equation}
N_{chains}=\frac{\phi_0V_0}{\nu v_{mon}} = \frac{\phi_0}{\nu}\left(\frac{R_0}{R_{mon}}\right)^3.
\end{equation}
where $v_{mon}=4\pi R_{mon}^3/3$. Hence, the elastic contribution, expressed in term of the polymer packing fraction (see eqn~(\ref{phi})), reads as
\begin{equation}
\label{Fm_elastic}
\beta F_m^{elastic}=\frac{3}{2}\frac{\phi_0}{\nu}\left(\frac{R_0}{R_{mon}}\right)^3\left[ \left(\frac{\phi}{\phi_0} \right)^{-2/3} +\frac{1}{3}\ln\left( \frac{\phi}{\phi_0}\right) -1 \right].
\end{equation}
We need to specify the packing fraction for the reference state, $\phi_0$. Most of the authors assume that such state is reached when the polymer is being cross-linked to create the permanent network. In other words, the reference state corresponds to the conformation of the particle when it was synthesized. In the particular case of microgels, the synthesis usually happens under bad solvent conditions, so the microgel is in the hydrophobic, collapsed state~\cite{Fern11}. Hence, we assume that $\phi_0 \approx 0.64$, which corresponds to the packing fraction in conditions of random-close packing of spheres (monomers).

The next step is to define the solvent free energy. In the typical Flory picture, this contribution has an entropic part and a solvent-polymer interaction part~\cite{Fern11,Fern00,Hoar07,Rumy15}
\begin{equation}
\beta F_m^{solvent} = n_s\ln \phi_s + \chi n_s \phi,
\end{equation}
where $n_s$ is the number of solvent molecules inside the microgel and $\phi_s$ is the volume fraction filled by the solvent inside the microgel, which is supposed to be given by $\phi_s\approx 1-\phi$. This assumption neglects the local packing fraction of absorbed counterions and should be valid for not too high microgel charges (about volumetric charge densities below $50$~C$/$cm$^3$). $\chi$ is the Flory-Huggins parameter (which controls the degree of solvent quality for the polymer chains).For $\chi=0$, polymers behave as athermal, so its conformation only depends on the excluded-volume interactions between the monomers. For $\chi=1/2$ the hydrophobic attraction between the polymer chains exactly compensates the excluded-volume repulsion, leading to an ideal Gaussian behavior. For larger values of $\chi$, the polymer tends to be more hydrophobic, so the microgel shrinks expelling the solvent from inside. The larger $\chi$ is, the smaller the microgel size becomes. For PNIPAM microgels $\chi$ increases with temperature, and the transition from swollen to shrunken conformations occurs at temperatures close to $T=307$~K. However, the dependence of $\chi$ with $T$ is not known in general, and depends on other parameters. For this reason we performed the study in terms of $\chi$ instead of using $T$. Assuming that the solvent particles have the same volume than the monomeric units, $v_{s}=v_{mon}$ we get
\begin{equation}
n_s=\frac{(1-\phi)V}{v_s}=(1-\phi) \left( \frac{R_m}{R_{mon}} \right)^3.
\end{equation}
Therefore,
\begin{equation}
\label{Fm_solvent}
\beta F_m^{solvent} = \left( \frac{R_m}{R_{mon}} \right)^3\left[ (1-\phi)\ln(1-\phi)-\chi \phi^2 \right]
\end{equation}
Please note that in the last equation we omitted a contribution proportional to $\chi\phi R_m^3 = \chi\phi_0 R_0^3$ because it only represents an additive constant that does not have any influence on the equilibrium swelling state.

Finally, the third contribution corresponds to the electrostatic energy of the charge distribution inside the microgel, in the absence of ions. Formally, this energy is given by
\begin{equation}
\beta F_m^{self-el}=\frac{l_B}{2}\int\int \frac{\rho_{mon}(\vec{r})\rho_{mon}(\vec{r}^\prime)}{|\vec{r}-\vec{r}^{\prime}|}d\vec{r}d\vec{r}^\prime
\end{equation}
In our particular model, we supposed a uniform distribution of charge inside the microgel, so that
\begin{equation}
\label{rhom}
\rho_{mon}(r)=\left\{\begin{array}{l@{\quad}l}
\frac{3Z_m}{4\pi R_m^3} & r\leq R_m \\
0 & r > R_m.
\end{array}  \right. 
\end{equation}
For this choice, the electrostatic self-energy of the microgel is simply given by
\begin{equation}
\label{Fm_selfel}
\beta F_m^{self-el}=\frac{3}{5}l_B\frac{Z_m^2}{R_m}.
\end{equation}
The total bare charge, $Z_m$, may be connected to the fraction of charged monomers, $f$ as
\begin{equation}
Z_m=-f\phi_0(R_0/R_{mon})^3.
\end{equation}
The negative sign is due to the fact that monomers are assumed to be negatively charged, $Z_{mon}=-1$. 

\subsection{Free energy of the ions}

The free energy of the ions is split into ideal and excess contributions 
\begin{equation}
F_{ions}[\{\rho_i(\vec{r})\}] = F_{ions}^{ideal}[\{\rho_i(\vec{r})\}] + F_{ions}^{excess}[\{\rho_i(\vec{r})\}], 
\end{equation}
for which the ideal gas part is known exactly as
\begin{equation}
\beta F_{ions}^{ideal}[\{\rho_i(\vec{r})\}]=\sum_{i=\pm}\int \rho_i(\vec{r})\left[ \ln ( \rho_i(\vec{r})\Lambda_i^3 ) -1\right]d\vec{r},
\end{equation}
where $\Lambda_i =h/(2\pi m_ik_BT)^{1/2}$ is the thermal wavelength of ion $i$. For the excess free energy of the ions, we assume that it may be written in two additive parts, a Coulombic electrostatic term plus a correction which accounts for the ionic correlations beyond the mean-field electrostatic contribution~\cite{Li04,Iyet82,Iyet83,Iyet84}. Therefore,
\begin{equation}
F_{ions}^{excess}[\{\rho_i(\vec{r})\}] = F_{ions}^{el}[\{\rho_i(\vec{r})\}] + F_{ions}^{corr}[\{\rho_i(\vec{r})\}]
\end{equation}
The mean-field electrostatic contribution is given by
\begin{eqnarray}
\label{coulomb}
\beta F_{ions}^{el}[\{\rho_i(\vec{r})\}] = \frac{l_B}{2}\sum_{i,j=\pm}Z_iZ_j\int\int\frac{\rho_i(\vec{r})\rho_j(\vec{r}^\prime)}{|\vec{r}-\vec{r}^{\prime}|}d\vec{r}d\vec{r}^{\prime} \nonumber \\
=\frac{l_B}{2}\sum_{i,j=\pm}Z_iZ_j\int\int\frac{( \rho_i(\vec{r})-\rho_i^b)(\rho_j(\vec{r}^\prime)-\rho_j^b)}{|\vec{r}-\vec{r}^{\prime}|}d\vec{r}d\vec{r}^{\prime}.
\end{eqnarray}
Note that the second equality holds because system fulfills electroneutrality (see eqn~(\ref{electroneutrality})). The correlation part may be estimated by means of different ways. Here, we employ one of the simplest approximations, in which this contribution is Taylor expanded up to second order of the local ionic concentrations around the bulk densities, and higher order terms are neglected:
\begin{eqnarray}
\label{fexcorr}
\beta F_{ions}^{corr}[\{\rho_i(\vec{r})\}]=\beta F_{ex}^{corr}[\{\rho_i^b\}]-\sum_{i=\pm}\Delta c_i^{(1)}\int (\rho_i(\vec{r})-\rho_i^b) d\vec{r} \nonumber \\
-\frac{1}{2}\sum_{i,j=\pm}\int\int \Delta c_{ij}^{(2)}(|\vec{r}-\vec{r}^\prime|) ( \rho_i(\vec{r}) - \rho_i^b )( \rho_j(\vec{r}^\prime) - \rho_j^b ) d\vec{r}d\vec{r}^{\prime}
\end{eqnarray}
In this expansion $\beta F_{ex}^{corr}[\{\rho_i^b\}]$, $\Delta c_i^{(1)}$ and $\Delta c_{ij}^{(2)}(|\vec{r}-\vec{r}^\prime|)$ are the Helmholtz free energy, the first-, and the second-order direct correlation functions of ions in the bulk (with uniform number density $\rho_i^b$) arising from the ionic correlations. As will be shown later on, $\beta F_{ex}^{corr}[\{\rho_i^b\}]$ and $\Delta c_i^{(1)}$ are two constants that don't have any influence on the ionic density profiles and the swelling ratio of the microgel at equilibrium. However, $\Delta c_{ij}^{(2)}(r)$ is indeed very important. Since the Coulomb contribution is explicitly taken into account through eqn~\ref{coulomb}, these functions are given by
\begin{equation}
\label{cij}
\Delta c_{ij}^{(2)}(r) = c_{ij}(r) + l_B\frac{Z_iZ_j}{r}  \ \ \ \ \ \ \ \ i,j=\pm
\end{equation}
where $c_{ij}(r)$ are the ion-ion direct correlation functions in the bulk.

\subsection{Functional differentiation}

Once all the contributions of the grand canonical potential are explicitly known, we can perform functional derivatives to obtain the ionic density profiles via minimization of the grand canonical potential
\begin{equation}
\frac{\delta \Omega\left[ \{ \rho_i(\vec{r}) \},R_m\right]}{\delta \rho_+(\vec{r})}=0 \ \ \ , \ \ \  \frac{\delta\Omega\left[ \{ \rho_i(\vec{r}) \}R_m\right]}{\delta \rho_-(\vec{r})}=0.
\end{equation}
Performing the functional differentiation, and using that $\rho_i(r)-\rho_i^b=\rho_i^bh_{mi}(r)$ together with the definition of the ion-ion correlation functions (eqn~\ref{cij}), we obtain
\begin{eqnarray}
\label{rhor}
\ln ( \rho_i(r)\Lambda_i^3 ) -\sum_{j=\pm}\rho_j^b\int c_{ij}(|\vec{r}-\vec{r}^\prime|)h_{mj}(\vec{r}^\prime)d\vec{r}^{\prime} \nonumber \\
+\beta V_{mi}(r) - \Delta c_i^{(1)}-\beta\mu_i  = 0  \ \ \ \ \ \ \ \ i=\pm
\end{eqnarray}
The same identity may be applied in the reservoir, where ions are uniformly distributed and so the density profiles are flat. In this case, there is no external field, so $h_{mi}(r)=0$, and the previous equation reduces to
\begin{equation}
\label{rhorbulk}
\ln ( \rho_i^b\Lambda_i^3 ) - \Delta c_i^{(1)}-\beta\mu_i = 0 \ \ \ \ \ \ \ \ i=\pm
\end{equation}
Using eqn~(\ref{rhorbulk}) in eqn~(\ref{rhor}), we can eliminate the constant parameters $\mu_i$, $\Delta c_i^{(1)}$ and $\Lambda_i$ in terms of the bulk densities $\rho_i^b$. Moreover, we can make use of the OZ equations (eqn~(\ref{OZ2})) to reduce even further. We obtain
\begin{eqnarray}
\label{rhor2bis}
\frac{\rho_i(r)}{\rho_i^b} &=& h_{mi}(r)+1 \\
&=&\exp \left[-\beta V_{mi}(r)+ h_{mi}(r)-c_{mi}(r) \right] \ \ \ \ \ \ \ \ i=\pm \nonumber
\end{eqnarray}
These expressions are completely consistent with the HNC formulation (eqn~(\ref{HNC})). Therefore, the DFT proposed here is equivalent to solving the OZ-HNC integral equations. Note that the present theory for the microgel-ion correlations reduces to the simpler mean-field Poisson-Boltzmann approach if ions are assumed to be point-like and ionic correlations are neglected ($\Delta c_{ij}^{(2)}=0$), so that the ion-ion direct correlation functions are provided by the mean spherical approximation (MSA), $c_{ij}= -l_BZ_iZ_j/r$.

Finally, the swelling state of the microgel is obtained by performing the derivative with respect to the particle radius, $R_m$, via
\begin{equation}
\frac{\delta \Omega\left[ \{ \rho_i(\vec{r}) \},R_m\right]}{\partial R_m} = 0 \ \ \ \ \ \ \ \ i=\pm.
\end{equation}

\section{Numerical implementation and choice of the conditions}
\label{implementation}

\begin{figure*}
 \centering
 \includegraphics[height=4.5cm]{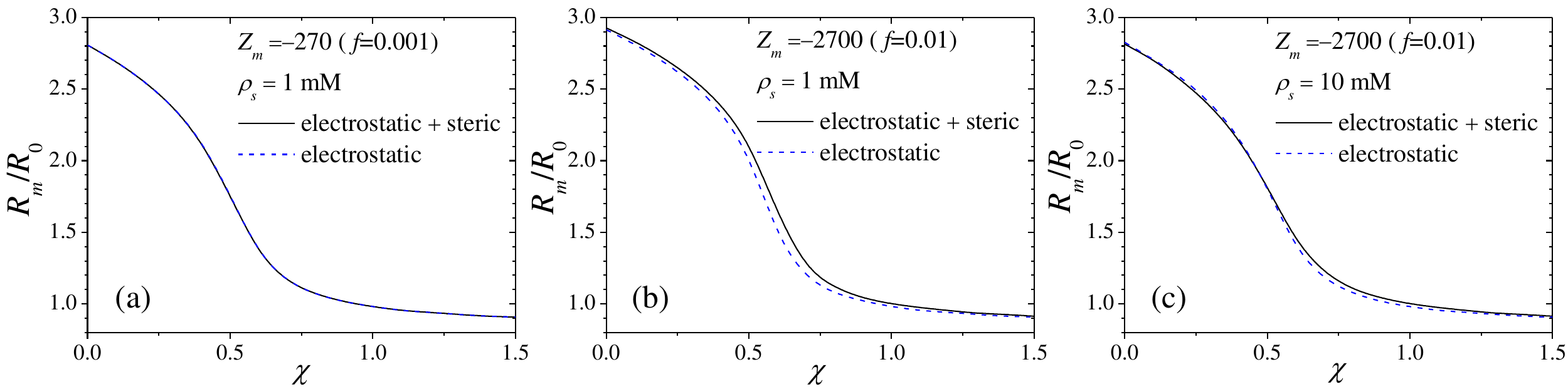}
 \caption{Swelling behavior of the microgel with and without the steric interaction. Each plot shows the curve for different conditions of charge fraction $f$ and electrolyte concentration $\rho_s$. Calculations performed for microgels with $\nu=500$ and $\phi_0=0.64$.}
 \label{compar_steric_electric}
\end{figure*}

In the calculations we investigate the particular case of monovalent salt, although the method can be also easily extended to consider multivalent ions. Also, we restrict the study to the case where the solvent mediated short-range hydrophobic/hydrophilic forces between ions and polymer chains are negligible, so that only steric and electrostatic forces are involved. The monomer radius is assumed to be $R_{mon}=0.4$~nm, which corresponds to the average effective size of the monomers for PNIPAM polymer chains. For the ions' diameter we employ a generic and symmetric $R_+=R_-=0.36$~nm, while in future more specific values~\cite{Kalcher1,Kalcher2} shall be tested. However, when an ion diffuses inside the cross-linked polymer network, the solvent layer surrounding the ion can be disrupted in the region between the ion and the polymer chain.\cite{Zhan07}. For instance, Na$^+$ is known to have some affinity to the carbonyl group of the PNIPAM amide bringing them to close contact~\cite{Alga11}. Still, the water disruption will depend on the specific ion and the detailed physicochemical properties of the polymer. In order to provide a simple treatment for this effect we assumed that the ion size which enters into the microgel-ion steric repulsion (eqn~\ref{vmister}) is the one for completely dehydrated ions and we fix it to $0.2$~nm in our calculations. The microgel radius in the reference (collapsed) state has been chosen to be $R_0=30$~nm, and the polymer packing fraction in this state is $\phi_0=0.64$, which corresponds to the random close packing configuration of spherical monomers. The Bjerrum length, that controls the intensity of the electrostatic interactions, is taken to be $l_B=0.71$~nm, which is the typical value for an aqueous solution at room temperature.

We investigate the equilibrium density profiles of counter- and coions, and the swelling state of the microgel at different conditions of microgel chain flexibility, bare charge and salinity. In particular, we explore charge fractions (average charge per monomer) from $f=0$ to $0.05$, whereas the salt concentration, $\rho_s$, is varied from 0.1~mM to 300~mM. The average chain length between two cross-linker nodes is varied between $\nu=100$ and $500$. In all cases electroneutrality is fulfilled. We also investigate the role of the steric repulsion between the incoming ions and the polymer network of the microgel. For this purpose the ionic density profiles and the swelling state of the particle are determined with and without taking the steric forces into account.

The hydrophobic interaction between the polymer chains is controlled by the Flory-Huggins parameter, $\chi$. It is well-known that the swelling of PNIPAM microgels depends on temperature, with a crossover between swollen to de-swollen states close to $T=307$~K. However, the dependence of $\chi$ with temperature is not universal, as it may be different depending on the polymer nature and the volume fraction. For this reason, instead of using the temperature we better employ the Flory-Huggins parameter, $\chi$. In order to plot the swelling curve, $\chi$ is varied from 0 (athermal polymer) to $1.5$ (strongly hydrophobic polymer), in steps of $\Delta \chi=0.1$.

In all the calculations the monomer diameter $\sigma_{mon}$ is set to be the unit length. All number densities are scaled by $\rho_i(r)\sigma_{mon}^3$. In order to integrate the OZ-HNC equations, successive iterations are applied starting from an initial guess until convergence is finally achieved. Iteration $k$ is considered to reach convergence when $\sum_{i<j}\int(c_{ij}^{(k)}(r)-c_{ij}^{(k-1)}(r))^2dr<10^{-9}$. Usually, as initial guess we employ the MSA approximation for the direct correlation function, $c_{ij}(r) = -\beta V_{ij}(r)$~\cite{Hans06}. The solution needs to be slowly conducted by mixing old and new iterations~\cite{Shah09}. This method works quite well for small values of the microgel bare charge, $Z_m$. However, for strongly charged microgels the method becomes unstable and the iterative procedure needs additional resources to be guided until convergence. In particular, we start from a microgel of large radius, since in this case the density distribution of bare charge is low and convergence is easily reached. Then, we use the functions $c_{m\pm}(r)$ obtained for this radius $R_m$ as the initial guess to solve the case of a microgel with slightly smaller size, $R_m-\Delta R$. For small enough $\Delta R$ the method is able to cover a large spectrum of particle sizes, from swollen to shrunken configurations, and allows the determination of the minimum of $\Omega$ as a function of $R_m$. This is done for several values of $\chi$ at once. The calculations were performed using a grid size of $r/\sigma_{mon}=0.005$, and a total number of points given by 262144. With such a choice we checked that the direct correlation functions are not affected by the grid size, even at the largest studied salt concentration. The step employed to cover the microgel radius, from large to small size, is $\Delta R = 0.2$~nm. This value ensures that the solution will successfully achieve convergence in all situations.  More details about the numerical integration of the OZ equations may be found in  related previous work~\cite{Monc13a}.

\section{Results and discussion}
\label{results}

\subsection{Effect of the microgel-ion excluded-volume repulsion}
\label{steric}

The first question to pose is whether the excluded-volume repulsion that the cross-linked polymer network exerts on the incoming ions is important or not. In principle, it is expected that this repulsion will have significant effects on the ionic distribution, expelling more ions outside the particle as the microgel shrinks. Analogously, the change on the ionic density profiles should also have some kind of influence on the swelling state of the particle. 

In order to answer this question the swelling behavior of the microgel has been studied under several conditions of bare charge and ionic strength (see Fig.~\ref{compar_steric_electric}). For weakly charge microgels, Fig.~\ref{compar_steric_electric}(a) indicates that the steric effect seems to be negligible. In this particular case, counterions are weakly attracted by the microgel, so the energy cost of expelling them outside due to the steric force is small compared to the other free energy contributions. However, the role of the steric exclusion becomes more relevant as the microgel bare charge is increased. This fact is clearly illustrated in Fig.~\ref{compar_steric_electric}(b), where the same comparison is performed for a microgel bare charge ten times larger. In this case, the steric exclusion induces an increase of the microgel size. Indeed, when steric exclusion is taken into account, counterions are forced to emigrate outside, increasing the electrostatic potential. Since this situation is energetically unfavorable, the microgel tends to swell in order to allow the ions to diffuse inside at some extent, and so reducing both the electrostatic and the steric microgel-ion energy. 

The relevance of the steric effect is not always the same for the whole curve. For $\chi < 1/2$ the particle is expanded so the internal polymer volume fraction is low, leading to a very small steric repulsion. As the particle shrinks ($\chi > 1/2$), the steric exclusion plays a more significant role, and may lead to an increase of the particle size up to 10\%. Close to the shrunken states the steric effect becomes very strong. However, in this region the microgel is highly hydrophobic so the free energy is dominated by the solvent contribution, being the presence of ions of minor relevance. The role that steric exclusion plays is also diminished by increasing the salt concentration. Indeed, if the salt concentration is large (see Fig.~\ref{compar_steric_electric}(c)) the electrostatic forces are screened and the presence of the microgel represents an smaller perturbation of the ionic bulk densities.

\begin{figure*}
 \centering
 \includegraphics[height=4.5cm]{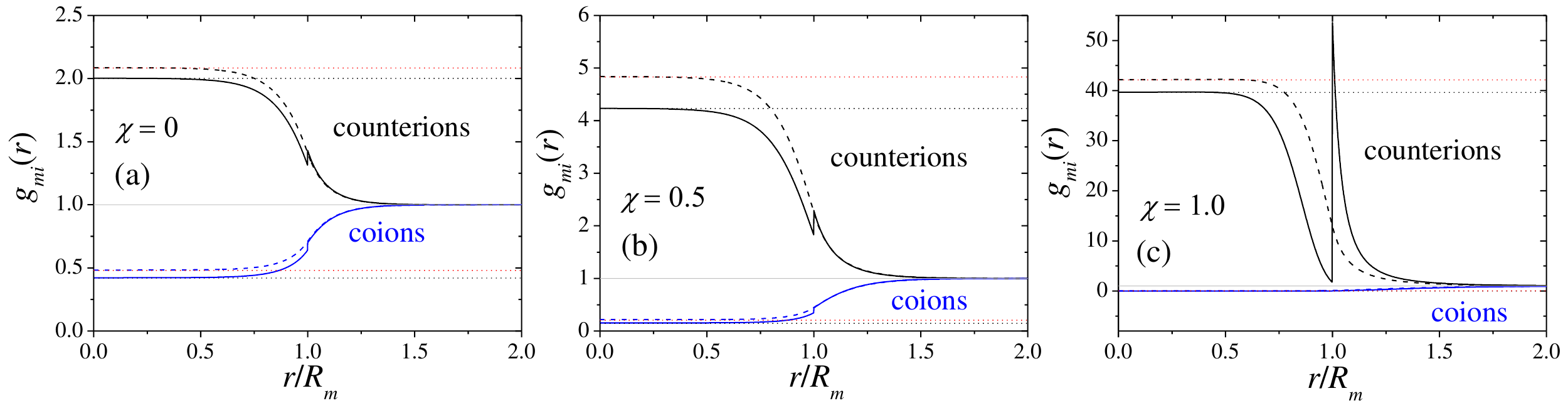}
 \caption{Radial distribution functions of counterions (+) a coions (-) around the microgel for three different swelling states, from (a) swollen to (c) shrunken. Solid and dashed lines are the predictions obtained with and without the steric exclusion, respectively. In all cases $f=0.01$ ($Z_m=-2700$), $\rho_s= 1$~mM, $\nu=500$ and $\phi_0=0.64$. Black and red dotted horizontal lines represent the ionic concentration predicted by eqns~(\ref{donnan})~and~(\ref{partition2}), providing a Donnan electrostatic potential with and without inclusion of the the excluded-volume effects, respectively.}
 \label{density_profiles_steric}
\end{figure*}
Although the microgel size is not very much affected by the steric exclusion, the ionic concentrations are indeed strongly modified. In order to illustrate this, Fig.~\ref{density_profiles_steric} shows an example of the ionic density profile with and without the steric effect for three different swelling states: swollen, intermediate and shrunken. As it may be observed in Fig.~\ref{density_profiles_steric}(a), in the swollen configuration ($\chi=0$) the effect of including the steric exclusion is small and only introduces a small jump in the radial distribution functions of both ions, in such a way that the concentration inside the microgel is slightly smaller than the one predicted without steric forces. When a partial shrinking is induced by increasing the Flory-Huggings parameter to $\chi=1/2$ (Fig.~\ref{density_profiles_steric}(b)), the differences between both predictions become more significant. Finally, close to the shrunken state (Fig.~\ref{density_profiles_steric}(c)) the effects of the steric exclusion are huge. In this case, the steric repulsive barrier is about several $k_BT$, causing a strong exclusion of ions. This can be clearly appreciated in the local concentration of counterions, which shows an accumulation peak at $r = R_m$ that grows at the expense of a reduction inside the microgel. Such kind of high concentration peaks located at the external shell of the microgel have been reported in previous simulation studies and are a direct consequence of the steric repulsion~\cite{Adro15}. The ion-ion excluded-volume repulsion also slightly contributes to enhance the counterion exclusion from the interior of the microgel. However, this effect becomes significant only when the internal  packing fraction of counterions is very large (close to the maximum packing) which can occur only for microgels holding a much higher density charge. Surprisingly, the enormous effects in the ionic distribution close to the shrunken state does not play a dominant role on the equilibrium swelling state. This occurs because, as it has been mentioned before, in this region of high $\chi$ the particle size is led by the polymer-polymer hydrophobic interactions which enters through the solvent term of the free energy.

In is important to emphasize that close to the microgel surface there exist local effects because of the sudden jump of the bare charge density. Indeed, the perturbation caused by the microgel interface leads to local variations of the local density of counter- and coions, with a range given approximately by the Debye length, $\kappa^{-1}$. However, for not too low salt concentrations ($\rho_s \gtrsim 0.1$~mM), the ionic density profiles become flat inside the microgel particle. This feature is directly connected with the fact that a Donnan potential ($\psi_D$) is being established inside the particle far from the interface, and that electroneutrality is fulfilled in this region. Such Donnan potential can not be determined from the original Donnan theory since it was developed only for point-like ions in the absence of excluded-volume interactions. As it was reported by Ahualli {\it et al.}~\cite{Ahua14}, the steric contribution must be necessarily taken into account, and the Donnan potential should be also corrected by finite size effects in addition to the electrostatic ones. They propose a simple theory to predict ionic concentration inside the gel phase by imposing the equality of the chemical potential inside (gel phase) and outside (bulk phase)~\cite{Ahua14}
\begin{equation}
\label{partition}
\rho_i^{ins}= \rho_i^b\exp \left( -\beta Z_ie\psi_D -\beta\mu_i^{exc} \right)
\end{equation}
where $\rho_i^{ins}$ and $\rho_i^b$ are the concentration of the ionic species $i=\pm$ inside the gel and in the bulk, $\psi^{D}\equiv\psi^{ins}-\psi^b$ is the difference of the electrostatic potential between both phases, and $\mu_i^{exc}$ is the excess chemical potential driven by the steric exclusion in the gel phase. By imposing electroneutrality inside the microgel, $Z_+\rho_+^{ins}+Z_+\rho_+^{ins}+Z_{mon}\rho_{mon}=0$, using that $Z_{mon}=-1$ and considering the particular case of a monovalent salt, $Z_+=+1$, $Z_-=-1$, $\rho_+^b=\rho_-^b=\rho_s$, the resulting Donnan potential is
\begin{equation}
\label{donnan}
\beta e\psi_D=-\beta \mu_+^{exc}  -\ln\left[ \frac{\rho_{mon}}{2\rho_s} +\sqrt{ \left(\frac{\rho_{mon}}{2\rho_s}\right)^2+ e^{-\beta(\mu_+^{exc}+\mu_-^{exc})}}\right].
\end{equation}

Now we need an expression for the excluded-volume chemical potentials $\mu_i^{exc}$. Here, we suppose that this steric repulsion is caused exclusively by the polymer packing fraction inside the microgel. This assumption neglects the excluded-volume effects arising from the counterions condensed inside, so it may be considered a good approximation as far as the microgel charge and the salt concentration are not too high. As it has been already mentioned in Section~\ref{system}, the internal morphology of the microgel is well-captured through the microgel-ion pair potential given by eqn~(\ref{vmister}). Therefore, in order to be consistent with this model, the excluded-volume chemical potential must be given by $\beta\mu_i^{exc} = -\ln (1-\phi)(1+R_i/R_{mon})^3$. Hence, the ionic densities deep inside the microgel particle are
\begin{equation}
\label{partition2}
\rho_i^{ins}= \rho_i^b\exp \left( -\beta Z_ie\psi_D +\ln(1-\phi)(1+R_i/R_{mon})^3 \right).
\end{equation}

The theoretical predictions privided by eqns~(\ref{donnan}) and~(\ref{partition2}) have been compared to the radial distribution functions of counter- and coions inside the microgel obtained in the region where the density profiles are flat (see black horizontal dotted lines in Fig.~\ref{density_profiles_steric}). As observed, the agreement is excellent for all cases, from swollen to shrunken states. Hence, eqn~(\ref{partition2}) can be considered a good approximation to predict the ionic concentration deep inside the microgel  in the presence of steric interactions. If the steric repulsion is neglected, the agreement is also good (red horizontal dotted lines), but in this case the Donnan potential is the one deduced from purely electrostatic interactions ($\mu_i^{exc}$=0). Furthermore, the fact that electroneutrality is fulfilled in the internal region means that the effective charge that the microgel develops must necessarily arise from the region close to the particle surface.

\begin{figure}[h]
\centering
  \includegraphics[height=6cm]{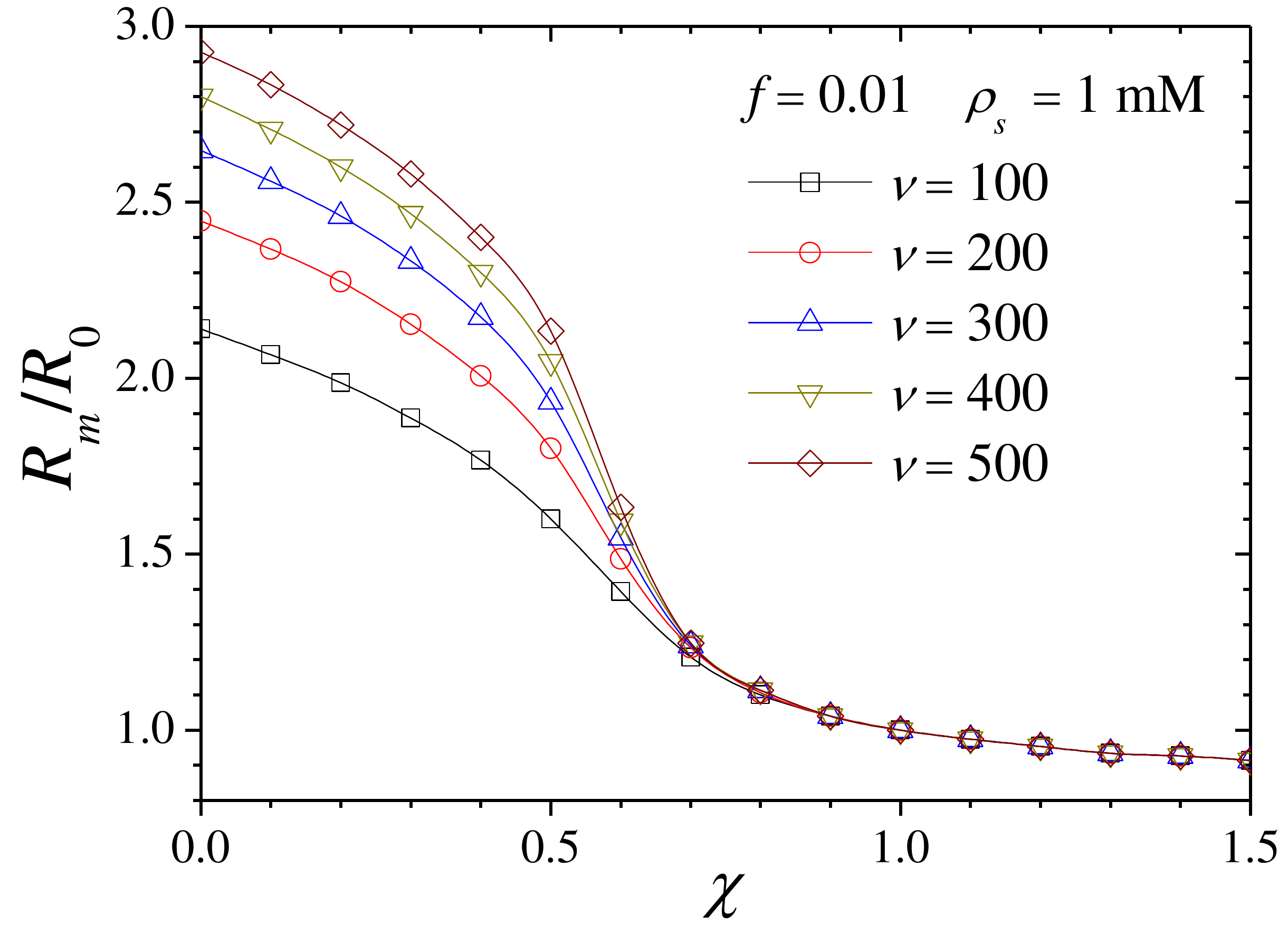}
  \caption{Swelling behavior for different values of the chain length between cross-linker nodes. Calculations performed for microgels with $\phi_0=0.64$, $f=0.01$ and $\rho_s= 1$~mM.}
  \label{Rm_Nu}
\end{figure}

\subsection{Effect of the chain flexibility}
\label{flex}

Henceforth, the steric exclusion will be always considered. We now investigate the role of the chain flexibility by changing the average chain length between two cross-linker monomers, $\nu$. Increasing the chain length is equivalent to decrease the cross-linker concentration inside the microgel particle, so the polymer network becomes more flexible. Fig.~\ref{Rm_Nu} shows the swelling behavior for different values of $\nu$, from 100 to 500. 
\begin{figure}[h]
\centering
  \includegraphics[height=12cm]{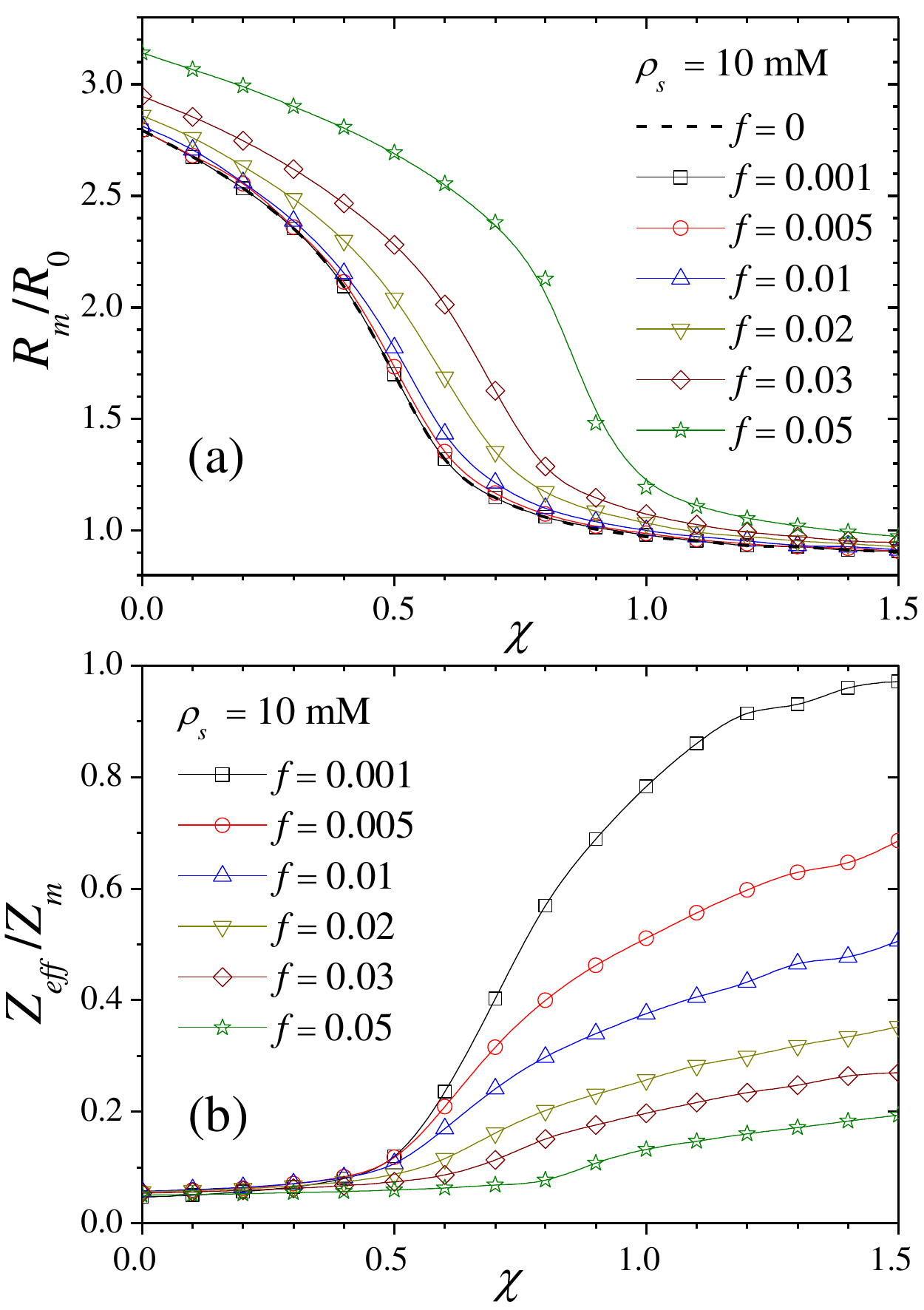}
  \caption{(a) Swelling behavior of the microgel for different bare charges, from $Z_m=0$ to $-13500$. (b) Effective charge of the microgel for different bare charges. Calculations were performed for microgels with $\nu=500$ and $\phi_0=0.64$ at a salt concentration of $\rho_s= 10$~mM.}
  \label{Rm_Zeff_chi}
\end{figure}

Clearly, the chain length does not have any relevant effect for shrunken states, where the polymer configuration is controlled by the solvent-induced polymer-polymer hydrophobic attraction. However, the swollen states are indeed strongly affected since the increase of $\nu$ allows the microgel to reach more expanded conformations with a lower free energy cost. These findings for the temperature dependence of the scaled microgel size are in very good qualitative agreement with experimental data for cross-linker densities between 0.6 and 5.3~\%~\cite{richtering}.

\subsection{Effect of the microgel bare charge}
\label{bare_charge}

\begin{figure*}
\centering
  \includegraphics[height=10.5cm]{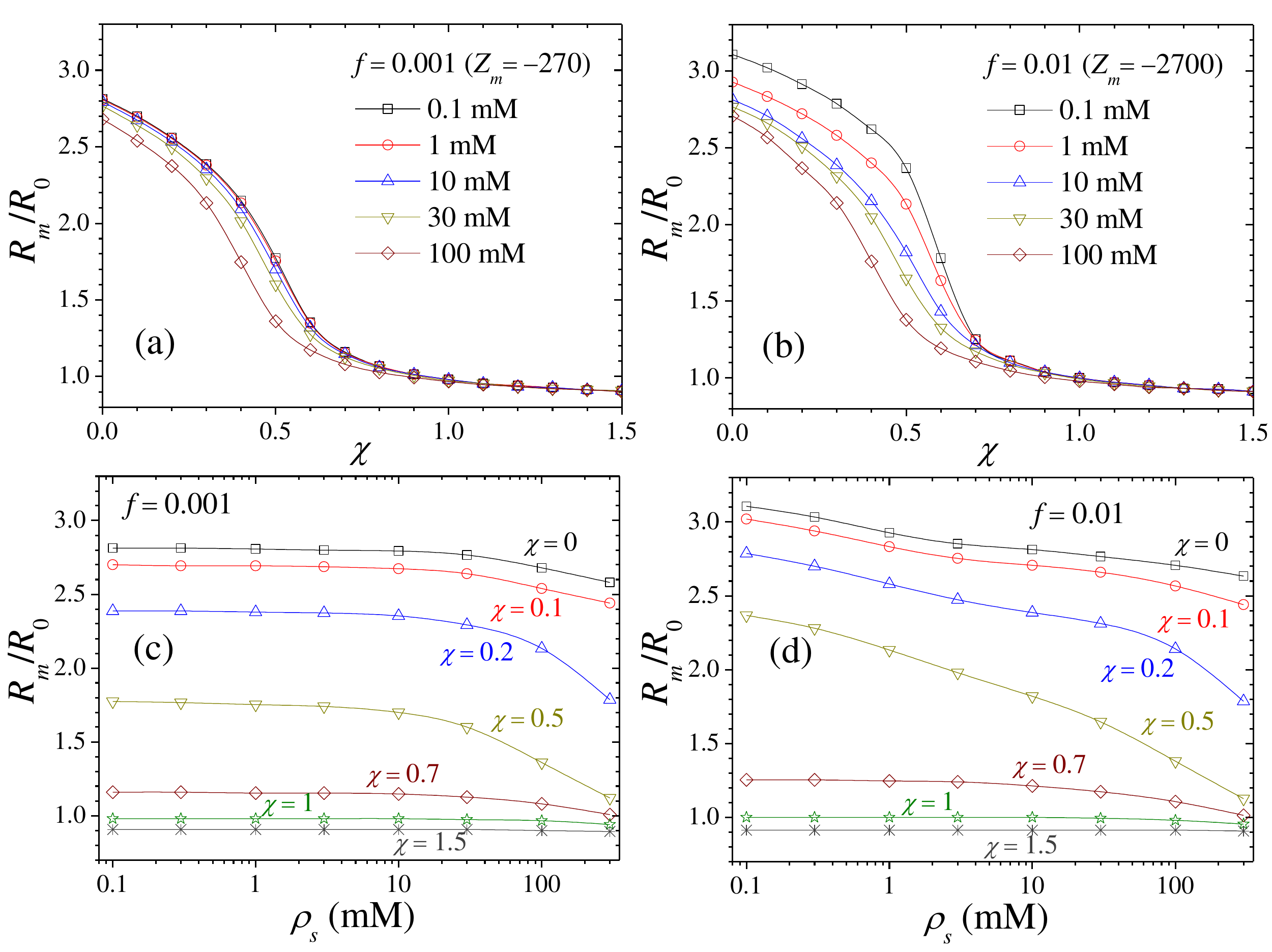}
  \caption{ Plots (a) and (b) show the normalized microgel radius versus $\chi$ for different salt concentrations for $f=0.001$ and $f=0.01$, respectively. Plots (c) and (d) show again the microgel size, but as a function of the salt concentration. Calculations were performed for microgels with $\nu=500$ and $\phi_0=0.64$.}
  \label{Rm_chi_salt}
\end{figure*}
The microgel bare charge, $Z_m$, has been varied from $0$ to $-13500$, which corresponds to a fraction of charged monomers ranging from $f=0$ to $0.05$. Fig.~\ref{Rm_Zeff_chi}(a) shows the swelling behavior for a fixed salt concentration given by $\rho_s= 10$~mM. As observed, for small bare charges the particle swelling is practically insensitive to $Z_m$, being dominated by the elastic and solvent-induced free energy contributions. However, this is not the case for large values of $Z_m$. Increasing $Z_m$ has essentially two effects: enlarging the particle size in the swollen state and shifting the volume transition to larger values of $\chi$. In other words, the microgel needs a stronger polymer-polymer hydrophobic attraction to compensate the electrostatic repulsion and induce the collapse. For $\chi$-values above the volume transition the swelling state in controlled by these hydrophobic forces and the results all converge to a common curve.

The ionic permeation inside the microgel is also affected by the value of the bare charge. This can be clearly appreciated in Fig.~\ref{Rm_Zeff_chi}(b), where the normalized effective charge, $Z_{eff}/Z_m$ calculated according to eqn~(\ref{effective_charge}) is plotted as a function of $\chi$ for different values of $Z_m$. In all cases, $Z_{eff}$ grows with $\chi$. This effect occurs for two reasons: First, increasing $\chi$ leads to an increase of the microgel volume fraction, which contributes to enhance the repulsive steric forces between counterions and the microgel. This counterion exclusion causes the increase of $Z_{eff}$. Second, a decrease of the microgel size forces the counterions to condensate inside a smaller volume. The counterion-counterion repulsion also contributes to the counterion exclusion and so, to the increase of the effective charge. It should be noted that the enhanced electrostatic repulsion between counterions arising in shrunken states is able to induce the increase of $Z_{eff}$ even in the absence of the steric exclusion effect (not shown). However, this increase is much less important than the one observed when considering the steric exclusion of the ions~\cite{Monc13b}.

Although $Z_{eff}$ always grows with $\chi$, the amount of increase is strongly dependent on the microgel bare charge, especially for de-swollen states. For small $Z_m$, the microgel-counterion electrostatic attraction is weak compared to the steric repulsive barrier at high values of $\chi$. This effect hinders the counterion permeation and yields a high effective charge, very close to the bare charge. However, for large values of $Z_m$, counterions become so strongly electrostatically attracted to the interior of the microgel that they are able to surpass the repulsive steric barrier and diffuse inside, inducing a significant reduction of $Z_{eff}/Z_m$. For swollen configurations ($\chi < 1/2$) counterions have plenty of space inside the microgel at the time that their permeation becomes mostly governed by electrostatics, so $Z_{eff}/Z_m$ tends to be independent on the bare charge. All these effects reveal an interesting interplay between electrostatic and excluded-volume effects.

\subsection{Effect of the salt concentration}
\label{salt_concentration}

The electrolyte concentration also plays an important role, as it is responsible for the screening of the electrostatic forces. Fig.~\ref{Rm_chi_salt}(a) plots the particle radius as a function of $\chi$ for increasing salt concentrations at a very low microgel bare charge ($f=0.001$). For such a weakly charged microgel, the response of the swelling ratio is almost insensitive to $\rho_s$ since the free energy is dominated by the elastic and solvent terms. When the particle charge increases (see Fig.~\ref{Rm_chi_salt}(b)), the eletrostatic term achieves a more relevant role and then, the swelling response becomes affected by the salt concentration. Basically, the effect of adding salt is to reduce the particle size in the swollen state and to shift the transition to lower $\chi$ values (temperatures). Both phenomena are consistent with previous simulation, theoretical, and experimental studies~\cite{Ques14b,Jiayin}. The explanation of this relies in the combination of several effects. On one hand, increasing the electrolyte concentration leads to the screening of the electrostatic forces, so we need smaller hydrophobic attraction to overcome the electrostatic repulsion and promote the particle shrinking. On the other hand, at high salt concentrations the ions outside the microgel particle generate an strong osmotic pressure on the microgel surface, which also favors the shrinking. Again, for large values of $\chi$, the swelling state becomes mostly controlled by the solvent-induced term, and all curves collapse in a common behavior.

Figs~\ref{Rm_chi_salt}(c) and (d) show the same sort of results, but plotted against $\rho_s$, for small and large bare charges. Again, it is clear that for weakly charged microgels, the swelling behavior is not affected by $\rho_s$. Indeed, only for a high enough concentration (above $\rho_s=10$~mM) there is an appreciable reduction of the particle size at low $\chi$. For strongly charged microgels the effect of the electrolyte concentration becomes more relevant and dominates over the solvent contribution. In this case, the particle de-swelling is observed even at $\rho_s=0.1$~mM. For both bare charges, the effect of the salt concentration is meaningless for shrunken microgels ($\chi>1$), where the particle size is almost entirely dominated by the hydrophobic interaction. Qualitatively similar plots may be found in the predictions of Colla {\em et al.}~\cite{Colla14}, obtained by solving the Poisson-Boltzmann equation. 

\begin{figure*}
\centering
  \includegraphics[height=10.5cm]{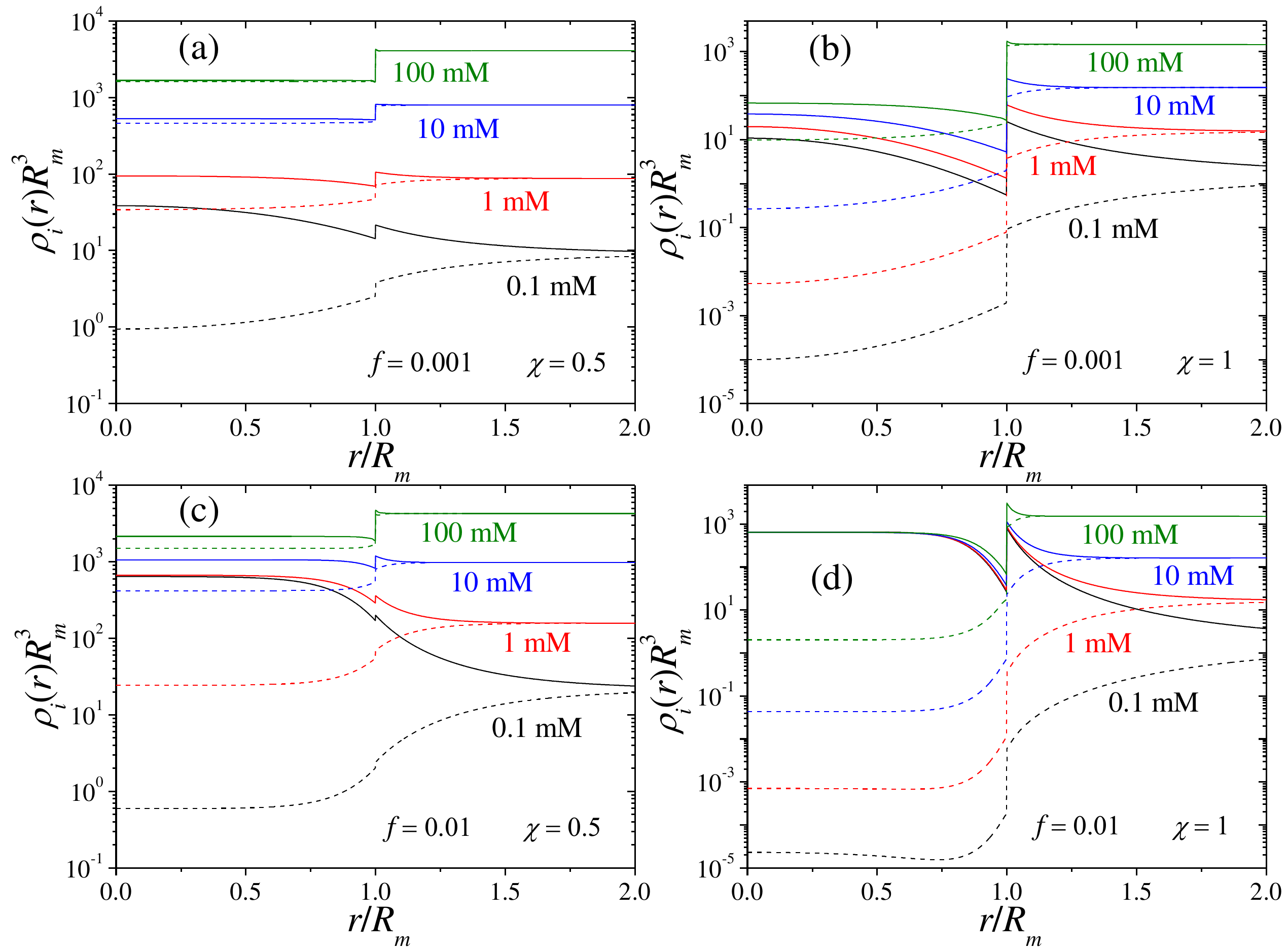}
  \caption{ Normalized density profiles of counterions (solid lines) and coions (dashed lines) at different salt concentrations for (a) $f=0.001$, $\chi=0.5$, (b) $f=0.001$, $\chi=1$, (c) $f=0.01$, $\chi=0.5$ and (d) $f=0.01$, $\chi=1$. Calculations were performed for microgels with $\nu=500$ and $\phi_0=0.64$.}
  \label{density_profiles_results}
\end{figure*}

In addition to the swelling state, our DFT also provides the equilibrium density profiles of counterions and coions inside and around the microgel particle. In Fig.~\ref{density_profiles_results} we plot these profiles for increasing values of the salt concentration, and for four different situations of bare charge and swelling states. We can extract some conclusions from these results. Firstly, the repulsive steric effect is present in all curves, leading to a certain reduction of the counterion and coion permeation. Evidently, this reduction is enhanced in the shrunken state, where the steric barrier is larger. For the case of counterions, the steric barrier is responsible for the appearance of an accumulation peak outside the microgel surface ($r \simeq R_m^+$) followed by a local minimum at the internal part of the surface $r\simeq R_m^-$. The height of the peak and the depth of the minimum become more pronounced when we move to shrunken states, since the steric jump becomes more important in this limit.

Secondly, increasing the electrolyte concentration flattens the ionic densities both outside and inside the microgel. Although for larger bare charges we need a larger amount of salt in order to homogenize the density profiles, in general these flat density profiles are obtained for a wide range of salt concentrations in the internal region of the microgel, far enough from the interface. As already mentioned before, these uniform profiles indicate that electroneutrality is satisfied in this region.

Thirdly, it is interesting to emphasize the results plotted in Fig.~\ref{density_profiles_results}(d), corresponding to the local density of ions for large $Z_m$ and $\chi=1$ (shrunken state). In this regime, the particle size is almost independent on $\rho_s$ (see Fig.~\ref{Rm_chi_salt}), so the steric repulsion is more or less the same in all cases. As a result of this, the counterion density profiles are very similar for all salt concentrations. This is not the case when we decrease the microgel bare charge (Fig.~\ref{density_profiles_results}(b)), as the counterion condensation inside the microgel grows with $\rho_s$. This phenomenon occurs because the steric repulsion plays here a more important role compared to the electrostatic attraction, and so it leads to a more pronounced minimum in the counterion density close to the microgel surface, which prevents at some extent the homogenization of the counterion density profile in the internal region.

\begin{figure*}
\centering
  \includegraphics[height=6cm]{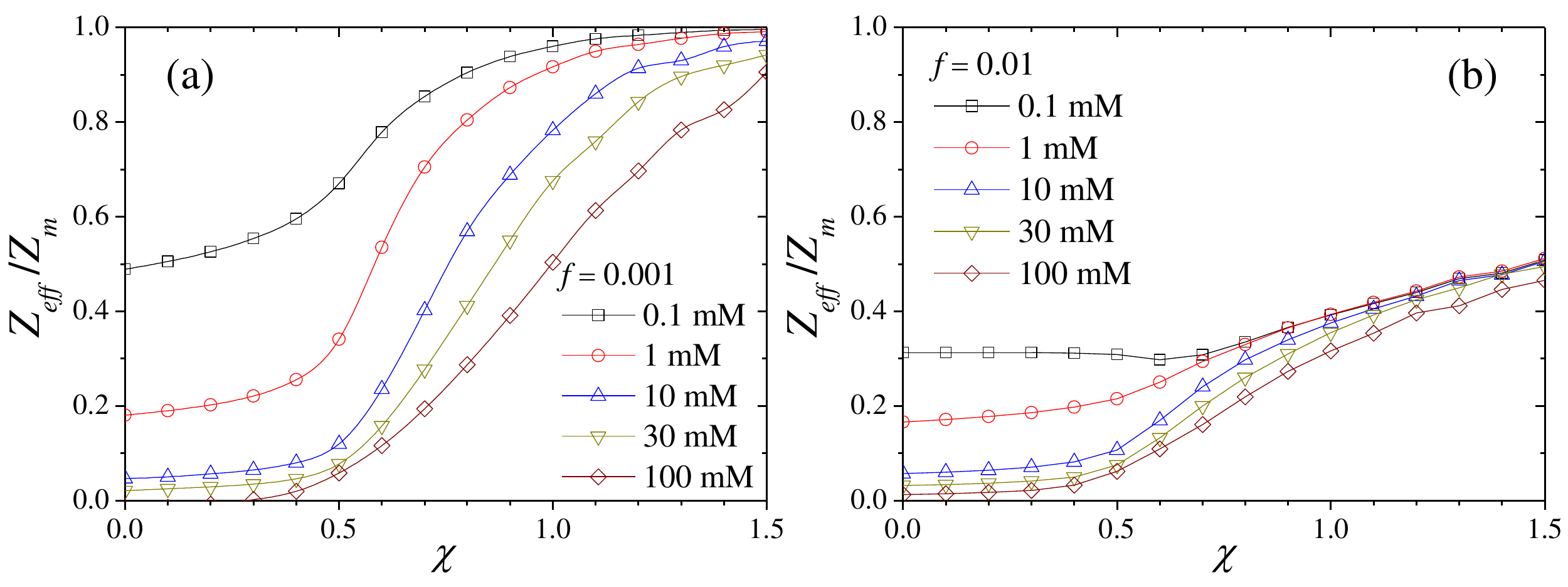}
  \caption{ Effective charge of the microgel as a function of $\chi$ at different salt concentrations, for (a) $f=0.001$ and (b) $f=0.01$. Calculations were performed for microgels with $\nu=500$ and $\phi_0=0.64$.}
  \label{Zeff_chi_salt}
\end{figure*}

As a global estimate of the charge screening provoked by the ionic double layer, we can examine the effective charge. The effect of the salt concentration on $Z_{eff}$ is shown in Figure~\ref{Zeff_chi_salt}. This parameter is much more sensitive to $\rho_s$ than the particle size. Indeed, $Z_{eff}$ may change even in situations where the particle radius is constant. The reason for such behavior relies on the fact that $Z_{eff}$ is controlled by the electrostatic interactions between ions and the microgel, whereas $R_m$ is also strongly influenced by the solvent-induced and elastic terms of the free energy. These last contributions mask the electrostatic effects, especially for small values of the microgel bare charge.

It is clear from both plots that increasing $\rho_s$ gives rise to a significant decay of the effective charge for swollen states ($\chi < 1/2$). Therefore, in general swollen microgels at moderate and high electrolyte concentrations are expected to hold very small effective charges compared to the bare ones. For small $Z_m$ (Fig.~\ref{Zeff_chi_salt}(a)) counterions are weakly attracted by electrostatics, but strongly repelled by steric exclusion in the shrunken state. In this regime, the steric forces are able to expel a large amount of counterions outside, leading to large effective charges, close to $Z_m$. For higher charged microgels (Fig.~\ref{Zeff_chi_salt}(b)), counterions are strongly attracted to the interior of the microgel by electrostatic forces, so they are able to surpass the steric exclusion leading to a a reduction of $Z_{eff}$. The decrease of $Z_{eff}$ with $\rho_s$ is more relevant in swollen configurations due to the screening of the electrostatic interactions. Surprisingly, for shrunken states the effect of adding salt is not significant, as all curves obtained for different salt concentrations merge for $\chi > 1$. The explanation of this phenomenon relies again in the fact that the counterion concentration inside the microgel is almost independent on the salinity under these conditions of bare charge and particle swelling (see again Fig.~\ref{density_profiles_results}(d)). In other words, for such strongly charged microgels counterions are forced to migrate inside in order to achieve electroneutrality. Hence, the steric barrier represents a small perturbation, only affecting the ionic concentration close to the interface. A similar conclusion has been reported by Colla {\em et al.}~\cite{Colla14} but using exclusively electrostatic interactions.

\section{Conclusions}
\label{conclusions}

The uniform swelling of microgels and the local distribution of 1:1 salt ions around the microgels have been studied by means of a new DFT framework. The theory accounts not only for the electrostatic interaction but also for the excluded-volume repulsive force that emerge when the ions diffuse inside the polymer network of the microgel. The free energy function is built in order to include the ion-ion correlations beyond the mean-field approach at the level of the HNC approximation.

The results show that the excluded-volume effect on the particle swelling is in general not very significant compared to the other free energy contributions such as the solvent-induced and the elastic terms, but becomes more important for strongly charged microgels and low electrolyte concentrations. The ionic density profiles, however, are deeply affected by the steric exclusion, especially for shrunken states. In particular, the steric repulsion enhances the accumulation of counterions at the external surface of the microgel and the subsequent reduction of its concentration inside the microgel particle compared to the predictions obtained with exclusively electrostatic interactions. This phenomenon can also be appreciated in the value of the microgel effective charge, $Z_{eff}$, which shows a very important enhancement as the microgel configuration goes from swollen to shrunken states.

We have found that the particle swelling is enhanced when increasing the microgel bare charge and/or decreasing the salt concentration, since both of them contribute to emphasize the electrostatic repulsion between the charged monomers inside the microgel. Moreover, the volume transition shifts to larger temperatures because additional hydrophobic polymer-polymer attractions are necessary to provoke the microgel shrinking. Analogously, increasing the chain flexibility (by enlarging the average chain length between cross-linker segments) also leads to more expanded configurations. For highly hydrophobic microgels ($\chi\gtrsim 1$) the microgel swelling is dominated by the solvent-polymer interaction so the particle size remains rather insensitive to all these parameters.

For a wide range of electrolyte concentrations, the ionic density profiles become uniform inside the microgel and only change near the particle interface due to the local effects caused by the polymer mass variation in there. Such a behavior clearly indicates that the internal part of the microgel is electroneutral, which means that the effective charge of the microgel arises in the region near the particle interface. Moreover, we have checked that the concentration of ions in the inner part of the microgel may be accounted for by means of a Donnan potential, but taking explicitly into account the excess chemical potential arising from the excluded-volume interaction. In other words, the ionic permeation is the result of a balance between electrostatic and steric effects. As a consequence of this, $Z_{eff}$ becomes almost insensitive to the salt concentration for shrunken microgels with large bare charge. In summary, the steric interaction should be always considered in realistic models to correctly predict the permeation of ions and charged or neutral solutes.

Our future research will focus on extending the present model to study the role of ion-specific hydrophobic/hydrophilic effects on microgel swelling, collapse transition~\cite{Zhan07,heyda:LCST}, and ionic permeation. In this respect, experimental, simulation and theoretical results with hydrophobic chaotropic ions have shown that the effective charge of the microgel is strongly sensitive to ion-specific absorption, leading to charge inversion and overcharging~\cite{Monc14,Lope07,Pere15}. Another interesting direction of investigation would be the inclusion of a fourth component such as a multipolar biomolecule, e.g., a globular protein, to study protein sorption equilibrium in hydrogels~\cite{yigit} or even the kinetics of protein sorption~\cite{stefano}. It would be also interesting to extend our model to study the non-homogeneous swelling of microgels caused by the non-uniform ionic density profiles. This could be done by means of a local free energy for the microgel that includes a position-dependent mass and charge distribution of the polymer network, just like the work by Rumyantsev{\em et al.}~\cite{Rumy15}. Finally, it is also worth to mention that the HNC-DFT connection established here is not only valid in the infinite dilute limit of microgels, but also in the so-called jellium approximation, where the macroion pair correlations are smeared out into a uniform background by setting $h_{mm}(r) = 0$~\cite{Triz04}. This identification opens up the possibility to investigate macroion properties in concentrated solutions with a degree of accuracy similar to one obtained from the traditional Wigner-Seitz (WS) cell approach~\cite{Colla10}. Under a qualitative point of view, for dense microgel suspensions the electric double layers strongly overlap, and the ionic concentrations are forced to be compressed in a smaller volume around the microgel in order to preserve electroneutrality. This effect would lead to a more abrupt change of the electrostatic potential at the microgel surface and so, to larger effectve charges.

\section*{Acknowledgements}

A.M-J. thanks the Spanish Ministerio de Econom\'{\i}a y Competitividad, Plan Nacional de Investigaci\'{o}n, Desarrollo e Innovaci\'{o}n Tecnol\'{o}gica (I+D+i, project MAT2012-36270-C04-02) and the European Regional Founding for financial support. J.D. acknowledges funding by the ERC (European Research Council) Consolidator Grant with project number 646659--NANOREACTOR.




\begin{thebibliography}{2}

\bibitem{Murr95}
M. J. Murray and M. Snowden,
{\em Adv. Colloid Interface Sci.\/}, 1995, {\bf 54}, 73--91.

\bibitem{Saun99}
B. R. Saunders and B. Vincent,
{\em Adv. Colloid Interface Sci.\/}, 1999, {\bf 80}, 1--25.

\bibitem{Fern11}
A. Fern\'{a}ndez-Nieves, H. M. Wyss, J. Mattsson and D. A. Weitz,
{\em Microgel Suspensions: Fundamentals and Applications} (Wiley-VCH, Weinheim, 2011).

\bibitem{Zhou15}
Y. Zhou, H. Tang and P. Wu,
{\em Phys. Chem. Chem. Phys.\/}, 2015, {\bf 17}, 25525--25535.

\bibitem{Tana79}
T. Tanaka and D. J. Fillmore,
{\em J. Chem. Phys.\/}, 1979, {\bf 70}, 1214--1218.

\bibitem{Long14}
G. S. Longo, M. Olvera de la Cruz and I. Szleifer,
{\em J. Chem. Phys.\/}, 2014, {\bf 141}, 124909.

\bibitem{Ramo11}
J. Ramos, A. Imaz, J. Callejas-Fern\'andez, L. Barbosa-Barros, J. Estelrich, M. Quesada-P\'erez and J. Forcada,
{\em Soft Matter\/}, 2011, {\bf 7}, 5067--5082.

\bibitem{Lesh}
S. C. Lesher-P\'{e}rez, T. Segura and C. Moraes
{\em Integr. Biol.\/}, 2016, {\bf 8}, 8--11.

\bibitem{Kenn16}
S. Kennedy, J. Hu, C. Kearney, H. Skaat, L. Gu, M. Gentili, H. Vandenburgh and D. Mooney,
{\em Biomaterials\/}, 2016, {\bf 75}, 91--101.

\bibitem{Levi02}
Y. Levin, A. Diehl, A. Fern\'{a}ndez-Nieves and A. Fern\'{a}ndez-Barbero,
{\em Phys. Rev. E\/}, 2002, {\bf 65}, 036143.

\bibitem{Lope04}
E. L\'{o}pez-Cabarcos, D. Mecerreyes, B. Sierra-Mart\'{\i}n, M. S. Romero-Cano, P. Strunz and A. Fern\'{a}ndez-Barbero,
{\em Phys. Chem. Chem. Phys.\/}, 2004, {\bf 6}, 1386--1400.

\bibitem{Dent03}
A. R. Denton,
{\em Phys. Rev. E\/}, 2003, {\bf 67}, 011804. Erratum {\em ibid}, 2003, {\bf 68} 049904.

\bibitem{Gott05} D. Gottwald, C. N. Likos, G. Kahl, and H. L\"owen,
{\em J. Chem. Phys.\/}, 2005, {\bf 122}, 074903.

\bibitem{Monc13a}
A. Moncho-Jord\'a, J. A. Anta and J. Callejas-Fern\'{a}ndez,
{\em J. Chem. Phys.\/}, 2013, {\bf 138}, 134902.

\bibitem{Monc13b}
A. Moncho-Jord\'a,
{\em J. Chem. Phys.\/}, 2013, {\bf 139}, 064906.

\bibitem{Monc14}
A. Moncho-Jord\'a and I. Adroher-Ben\'{\i}tez,
{\em Soft Matter\/}, 2014, {\bf 10}, 5810.

\bibitem{Fern00}
A. Fern\'{a}ndez-Nieves, A. Fern\'{a}ndez-Barbero, B. Vincent and F. J. de las Nieves,
{\em Macromolecules\/}, 2000, {\bf 33}, 2114--2118.

\bibitem{Hoar07}
T. Hoare and R. Pelton,
{\em J. Phys. Chem. B\/}, 2007, {\bf 111}, 11895--11906.

\bibitem{Koso15}
P. Kosovan, T. Richter and C. Holm, 
{\em Macromolecules\/}, 2015, {\bf 48}, 7698--7708.

\bibitem{Rumy15}
A. E. Rumyantsev, A. A. Rudov and I. I. Potemkin,
{\em J. Chem. Phys.\/}, 2015, {\bf 142}, 171105.

\bibitem{Rumy14}
A. M. Rumyantsev, S. Santer and E. Y. Kramarenko,
{\em Macromolecules\/}, 2014, {\bf 47}, 5388--5399.

\bibitem{Sing13}
C. E. Sing, J. W. Zwanikken and M. Olvera de la Cruz, 
{\em Macromolecules\/}, 2013, {\bf 46}, 5053--5065.

\bibitem{Ahua14}
S. Ahualli, A. Mart\'{\i}n-Molina and M. Quesada-P\'erez,
{\em Phys. Chem. Chem. Phys.\/}, 2014, {\bf 16}, 25483--25491.

\bibitem{Colla14}
T. Colla, C. N. Likos and Y. Levin
{\em J. Chem. Phys.\/}, 2014, {\bf 141}, 234902.

\bibitem{Adro15}
I. Adroher-Ben\'{\i}tez, S. Ahualli, A. Mart\'{\i}n-Molina, M. Quesada-P\'erez and A. Moncho-Jord\'a,
{\em Macromolecules\/}, 2015, {\bf 48}, 4645--4656.

\bibitem{Jian07}
T. Jiang, Z. Li and J. Wu,
{\em Macromolecules\/}, 2007, {\bf 40}, 334-343.

\bibitem{Jian08}
T. Jiang and J. Wu,
{\em J. Phys. Chem. B\/}, 2008, {\bf 112}, 7713--7720.

\bibitem{Li06}
Z. Li and J. Wu,
{\em Phys. Rev. Lett.\/}, 2006, {\bf 96}, 048302.

\bibitem{Alts87}
T. Alts, P. Niebala, B. D'Aguanno and F. Forstmann,
{\em Chemical Physics\/}, 1987, {\bf 111}, 223--240.

\bibitem{Li04}
Z. Li and J. Wu,
{\em Phys. Rev. E\/}, 2004, {\bf 70}, 031109.

\bibitem{Yu04}
Y-X. Yu, J. Wu and G-H. Gao,
{\em J. Chem. Phys.\/}, 2004, {\bf 120}, 7223-7233.

\bibitem{Fati03}
N. Fatin-Rouge, A. Milon, J. Buffle, R. R. Goulet and A. Tessier
{\em J. Phys. Chem. B\/}, 2003, {\bf 107}, 12126--12137.

\bibitem{Ques14a}
M. Quesada-P\'erez, I. Adroher-Ben\'itez and J. A. Maroto-Centeno,
{\em J. Chem. Phys\/}, 2014, {\bf 140}, 204901.

\bibitem{Lazz00}
M. J. Lazzara, D. Blankschtein and W. M. Deen,
{\em J. Colloid Interface Sci.\/}, 2000, {\bf 226}, 112--122.

\bibitem{Hans06}
J. P. Hansen and I. R. McDonald,
{\em Theory of Simple Liquids\/} 3rd Ed. (Academic Press, USA, 2006).

\bibitem{Iyet82}
H. Iyetomi and S Ichimaru,
{\em Phys. Rev. A\/}, 1982, {\bf 25}, 2434--2436.

\bibitem{Iyet83}
H. Iyetomi and S Ichimaru,
{\em Phys. Rev. A\/}, 1983, {\bf 27}, 3241--3250.

\bibitem{Iyet84}
H. Iyetomi,
{\em Progress of Theoretical Physics\/}, 1984, {\bf 71}, 427--437.

\bibitem{Kalcher1}
I. Kalcher, J. C. F. Schulz and J. Dzubiella
{\em Phys. Rev. Lett.\/}, 2010, {\bf 104}, 097802.

\bibitem{Kalcher2}
I. Kalcher, J. C. F. Schulz and J. Dzubiella
{\em J. Chem. Phys.\/}, 2010, {\bf 132}, 164511.

\bibitem{Zhan07}
Y. Zhang, S. Furyk, L. B. Sagle, Y. Cho, D. E. Bergbreiter and P. S. Cremer,
{\em J. Phys. Chem. C\/}, 2007, {\bf 111}, 8916--8924.

\bibitem{Alga11}
E. A. Algaer and N. F. A. van der Vegt,
{\em J. Phys. Chem. B\/}, 2011, {\bf 115}, 13781--13787

\bibitem{Shah09}
N. H. Shah,
{\em Numerical Methods with C++ programming\/} (PHI, 2009).

\bibitem{Ques14b}
M. Quesada-P\'erez, S. Ahualli and A. Mart\'{\i}n-Molina
{\em J. Chem. Phys\/}, 2014, {\bf 141}, 124903.

\bibitem{Lope07}
T. L\'opez-Le\'on, A. Ala\"issari, J. L. Ortega-Vinuesa, and D. Bastos-Gonz\'alez,
{\em ChemPhysChem\/}, 2007, {\bf 8}, 148--156.

\bibitem{Pere15}
L. P\'{e}rez-Fuentes, C. Drummond, J. Faraudo and D. Bastos-Gonz\'{a}lez,
{\em Soft Matter\/}, 2015, {\bf 11}, 5077--5086.

\bibitem{richtering}
H. Senff and W. Richtering, 
{\em Colloid. Polym. Sci.\/}, 2000, {\bf 278}, 830--840.

\bibitem{Jiayin}
J. Heyda, S. Soll, J. Yuan and J. Dzubiella,
{\em Macromolecules\/}, 2014, {\bf 47}, 2096--2102.

\bibitem{heyda:LCST}
J. Heyda and J. Dzubiella,
{\em J. Phys. Chem B.\/}, 2014, {\bf 118}, 10979--10988.

\bibitem{yigit}
C. Yigit, N. Welsch, M. Ballauff and J. Dzubiella, 
{\em Langmuir\/}, 2012, {\bf 28}, 14373--14385.

\bibitem{stefano}
S. Angioletti-Uberti, M. Ballauff and J. Dzubiella, 
{\em Soft Matter\/}, 2014, {\bf 10}, 7932--7945.

\bibitem{Triz04}
E. Trizac and Y. Levin, 
{\em Phys. Rev. E\/}, 2004, {\bf 69}, 031403.

\bibitem{Colla10}
T. E. Colla and Y. Levin, 
{\em J. Chem. Phys.\/}, 2010, {\bf 133}, 234105.

\end{thebibliography}

\end{document}